\documentclass[11pt]{article}

\usepackage[preprint]{acl}
\usepackage{amsmath, bm} 
\usepackage{times}
\usepackage{latexsym}
\usepackage{booktabs}
\usepackage{array}
\usepackage{multirow}
\usepackage{colortbl}
\usepackage{xcolor}
\usepackage{booktabs}

\usepackage[T1]{fontenc}

\usepackage[utf8]{inputenc}

\usepackage{microtype}

\usepackage{inconsolata}

\usepackage{graphicx}

\title{EvoConfig: Self-Evolving Multi-Agent Systems for Efficient Autonomous Environment Configuration}

\author{
Xinshuai Guo$^{1}$\thanks{\ \ indicates equal contribution.},~Jiayi Kuang$^{2*}$,~Linyue Pan$^{1}$,~Yinghui Li$^{3}$\thanks{\ \ Corresponding authors.},~Yangning Li$^{1}$\\~\textbf{Hai-Tao Zheng}$^{1\dagger}$,~\textbf{Ying Shen}$^{2\dagger}$,~\textbf{Di Yin}$^{3}$,~\textbf{Xing Sun}$^{3}$\\
        $^{1}$ Tsinghua University, $^{2}$ Sun-Yat Sen University, $^{3}$ Tencent Youtu Lab \\
        \texttt{gxs25@mails.tsinghua.edu.cn}, \texttt{liyinghuihhh@gmail.com}
}

\begin{document}
\maketitle
\begin{abstract}
A reliable executable environment is the foundation for ensuring that large language models solve software engineering tasks. 
Due to the complex and tedious construction process, large-scale configuration is relatively inefficient. However, most methods always overlook fine-grained analysis of the actions performed by the agent, making it difficult to handle complex errors and resulting in configuration failures. 
To address this bottleneck, we propose EvoConfig, an efficient environment configuration framework that optimizes multi-agent collaboration to build correct runtime environments. EvoConfig features an expert diagnosis module for fine-grained post-execution analysis, and a self-evolving mechanism that lets expert agents self-feedback and dynamically adjust error-fixing priorities in real time. 
Empirically, EvoConfig matches the previous state-of-the-art Repo2Run on Repo2Run’s 420 repositories, while delivering clear gains on harder cases: on the more challenging Envbench, EvoConfig achieves a 78.1\% success rate, outperforming Repo2Run by 7.1\%. Beyond end-to-end success, EvoConfig also demonstrates stronger debugging competence, achieving higher accuracy in error identification and producing more effective repair recommendations than existing methods~\footnote{We will open-source the code after the paper is published.}.
\end{abstract}

\begin{figure}[t]
    \centering
    \includegraphics[width=0.5\textwidth]{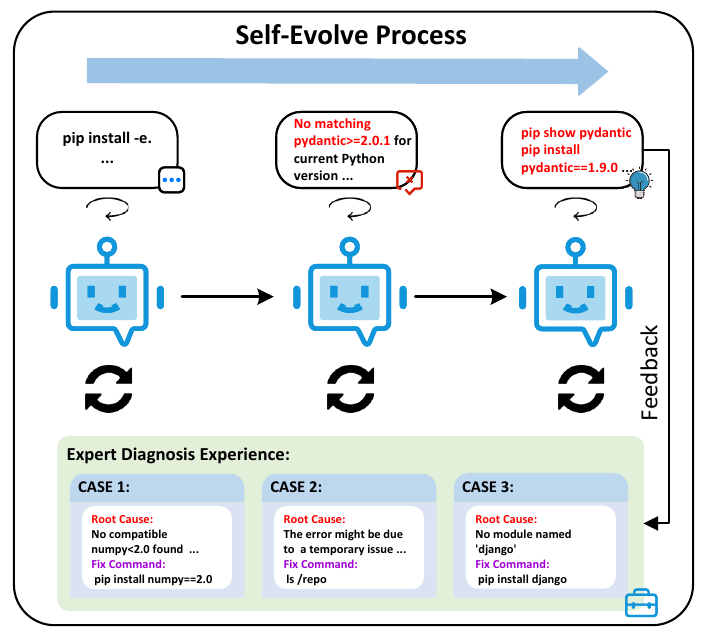}
    \caption{Self-Evolving Diagnostic Process.}
    \label{fig:self-evo}
\end{figure}

\section{Introduction}

Large language models (LLMs) have made rapid progress in handling complex software engineering (SWE) tasks~\citep{he2025llm, wang2025agents, xia2025live, wang2024opendevin, kuang2025atomic, lu2025youtu}, leading to the emergence of a wide range of code agents such as SWE-Agent~\citep{yang2024swe}, OpenHands~\citep{wang2024openhands},  MetaGPT~\citep{hong2023metagpt}, Copilot~\citep{githubcopilot} and Cursor~\citep{cursor}. As research increasingly shifts toward repository-level software engineering tasks, scalable execution and reliable validation become essential~\citep{wang2025mirror, xie2024codebenchgen, liu2024scaling, rutherford2024jaxmarl, krnjaic2024scalable}. Code agents are no longer required to only generate or modify code, but must also complete end-to-end workflows, including environment construction, testing, and validation, within real code runtime environments. However, a long-overlooked challenge is now becoming increasingly evident: automatically configuring executable environments. Environment setup still depends heavily on human expertise and can be difficult even for experienced developers. Yet a stable, runnable environment is a prerequisite for tackling complex software engineering tasks. \textit{Therefore, enabling agents to reliably configure environments is critical to advancing code agents}.

In real-world repositories, an agent must autonomously complete dependency installation, version resolution, and test execution under challenging conditions, including unknown dependencies, incomplete documentation, and the coexistence of multiple build tools. Most existing methods formulate environment configuration as a sequential decision-making problem~\citep{bouzenia2025you, vergopoulos2025automated, hu2025reporun, zhang2025autoenv}: the agent observes the current execution outcome and heuristically proposes the next action. However, \textbf{these approaches often fail to explicitly address process-level errors that occur during configuration} (e.g., cascading dependency conflicts, toolchain mismatches, partial installations). Such errors can accumulate across steps and ultimately cause environment construction to fail.
Some prior work~\citep{milliken2025beyond, vergopoulos2025automated} introduces repair strategies that are detached from the original configuration context. These methods typically rely on predefined, experience-based rules to produce static repair actions, \textbf{but they lack fine-grained diagnosis of the specific failure causes in the ongoing configuration process}. Consequently, agents are more prone to hallucinated fixes or repetitive trial-and-error behaviors, and may even fall into infinite loops when confronted with complex failures. In addition, the single-agent workflow exacerbates the problem: error-related information and noisy execution traces accumulate over time, which can mislead subsequent decisions and further reduce both success rate and efficiency.

To address these challenges, we propose EvoConfig, an efficient environment configuration framework based on self-evolving multi-agent collaboration. \textbf{Our core objective is to improve environment configuration success rates while simultaneously enhancing process-level error correction capabilities during the configuration process}. Specifically, a main agent is responsible for environment configuration, while expert agents act as diagnostic specialists that perform fine-grained analysis of execution results and autonomously determine whether repairs are required, ultimately providing structured and actionable guidance to the main agent. More importantly, we introduce an online self-evolving mechanism that enables expert agents to continuously learn from error correction cases and dynamically adjust their analytical focus and structured suggestions, thereby improving the agent’s ability to resolve complex environment configuration failures. Notably, this self-evolving mechanism does not rely on external memory modules, avoiding additional reasoning overhead and token consumption. We evaluate EvoConfig on multiple real-world open-source repositories, and the results demonstrate that our approach not only improves environment configuration success rates but also significantly enhances process-level error correction during the configuration process.

In summary, our main contributions are summarized as follows:
\begin{itemize}
    \item We are the first to propose the multi-agent collaborative framework EvoConfig for automated environment configuration, improving configuration success rates through optimized agent workflows.
    \item We propose an expert diagnostic module and introduce a self-evolving mechanism to adaptively enhance the process-level error correction capability of agents in the environment configuration process.
    \item We conduct extensive evaluations on multiple open-source benchmarks against advanced agent frameworks, demonstrating that EvoConfig achieves state-of-the-art performance in both environment configuration and process-level error correction.
\end{itemize}

\section{Formulation}

\subsection{Task Definition}

Given a real-world open-source GitHub repository $R$ at a specified version, the system is provided with a clean initial execution environment $E_0$. The ultimate goal is to automatically construct a target execution environment $E$ through a sequence of interactive commands, such that unit tests can be successfully executed in the resulting environment.

\subsection{Iterative Configuration Process}

In this work, we model environment configuration as an interactive decision-making process. Specifically, at interaction round $t$, the agent is in the current environment state $E_t$ and selects a set of commands from the action space $\mathcal{A}$ for execution:
\begin{equation}
a_t = \{c_t^1, c_t^2, \dots, c_t^{k_t}\}, \quad a_t \subseteq \mathcal{A},
\end{equation}
where each $c_t^i$ denotes an atomic executable command, and $k_t$ is the number of commands issued at round $t$.

After executing the command set $a_t$, the system performs a state transition based on the current environment state and the execution outcomes:
\begin{equation}
E_{t+1} = \delta(E_t, a_t).
\end{equation}

This process is repeated for at most $t_{\max}$ interaction rounds, until the test cases are successfully executed or the number of interactions exceeds a predefined maximum threshold. In this work, we place particular emphasis on whether the commands generated at each round result in execution errors. Accordingly, the overall optimization objective is to improve the environment configuration success rate under a limited interaction budget, while simultaneously enhancing the agent’s capability for process-level error correction.

\begin{figure*}[t]
    \centering
    \includegraphics[width=1.23\textwidth]{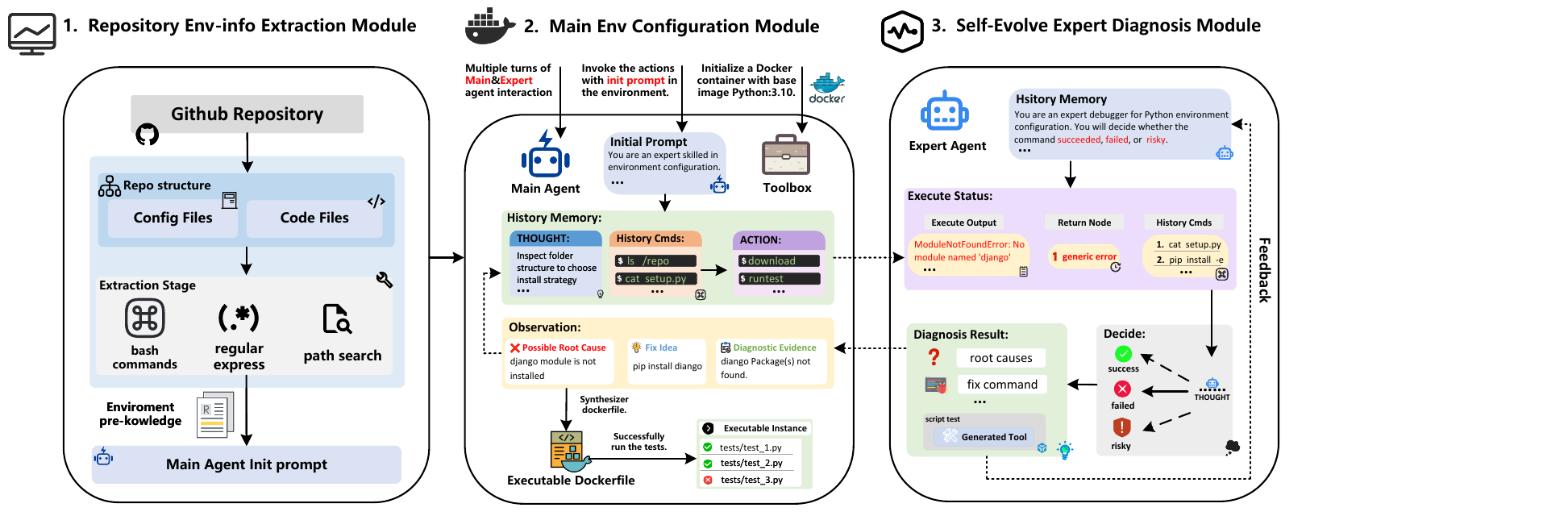}
    \caption{Workflow of EvoConfig. A \textbf{main configuration agent} performs interactive environment setup, while \textbf{self-evolving expert diagnostic agent} analyzes execution feedback and provide adaptive guidance. The validated command sequence is consolidated into a runnable Dockerfile.}

    \label{fig:evoconfig}
\end{figure*}

\section{Method}

This section introduces EvoConfig, a self-evolving multi-agent framework for efficient environment configuration. Given a code repository, EvoConfig performs multiple rounds of interaction and decision-making while continuously repairing environment configuration issues, ultimately generating an executable Dockerfile to build a runnable environment. EvoConfig consists of three main components: an environment information extraction module, a main environment configuration module, and an self-evolving expert diagnosis module.

\subsection{Environment Info Extraction Module}

We introduce a lightweight environment information extraction module that provides the main agent with a small set of high-impact prior signals before interactive configuration begins. The module focuses on extracting stable structural cues that are directly relevant to environment configuration.

Formally, given a repository $R$, the module produces a prior summary:
\begin{equation}
  P(R) = \{M, I, T\},
\end{equation}
where $M$, $I$, and $T$ denote the dependency management strategy, project importability, and test structure, respectively.

\paragraph{Dependency Management Strategy.} The dependency management strategy $M$ is inferred from configuration files such as \texttt{poetry.lock}, \texttt{pyproject.toml}, and \texttt{requirements*.txt}, guiding early installation decisions. 
\paragraph{Project Importability.} Project importability $I$ captures whether the project needs to be installed for tests to run, based on installation metadata, \texttt{src/} layouts, and package structure. 
\paragraph{Test Structure Hypothesis.} The test structure hypothesis $T$ describes the presence and location of tests, the inferred test framework, and whether tests import project modules.

The prior summary $P(R)$ is injected into the initial prompt words of the main agent to guide the generation of the initial configuration strategy of the main agent with almost no increase in computational cost.

\subsection{Main Environment Configuration Module}

After the environment prior information is extracted, the system enters the core environment configuration stage. Unlike previous approaches, the main agent responsible for environment configuration focuses solely on action execution and sequence, without bearing the burden of long-term memory and semantic analysis of execution results.

Specifically, at interaction step $t$, the main agent performs ReAct~\citep{yao2022react} framework reasoning based on a limited context and generates an action output, which is parsed into a sequence of atomic commands and executed sequentially in the runtime environment. Each command returns standard output and an exit code as execution feedback for the current step. During this process, the main agent concentrates only on action generation, scheduling, and execution order, and does not directly interpret the semantics of execution results. Instead, the execution context is delegated to the expert diagnosis module for analysis. This design allows the main agent to advance execution in a streaming manner, avoiding the accumulation of large volumes of raw output across multiple interaction rounds and receiving only highly summarized analytical feedback. As a result, it effectively mitigates a key issue in traditional interactive systems, where incorporating large amounts of low-value output directly into the main reasoning context leads to memory inflation and interferes with subsequent decision-making.

In addition, to maintain reasoning quality while reducing overall overhead, the main agent adopts a strict context management strategy. Combined with the system’s rollback mechanism, it preserves key command sequences from successful execution rounds as well as structured diagnostic summaries from the diagnosis module as experience, thereby improving the efficiency of action generation and scheduling during environment configuration.

\subsection{Self-Evolving Expert Diagnosis Module}

During environment configuration, accurate error diagnosis and repair are critical to ensuring successful system deployment. To enhance process-level error correction capability, we introduce an expert diagnosis module with a self-evolving mechanism, which explicitly decouples the standard output of execution actions from the primary configuration workflow and assigns it to an independent expert diagnostic agent.

The critical function of the expert agent is to assess the outcomes of execution actions and produce fine-grained analytical results. Specifically, based on the executed command, exit code, and standard output, the expert agent categorizes each action into one of three states—\textbf{success}, \textbf{failure}, or \textbf{potential risk}. According to the identified state, it generates corresponding repair commands or risk suggestions, and ultimately outputs a structured diagnostic report.

Notably, the expert agent is endowed with the capability of \textbf{on-the-live tool} creation and execution. At each decision step, the agent autonomously determines whether auxiliary tools are needed to support error judgment. Tool creation is subject to strict constraints: each tool must be a \textbf{single-line executable command} used solely for collecting diagnostic evidence rather than performing repairs. The outputs of these tools are treated as diagnostic evidence to strengthen error interpretation and are fed back to the main agent in a structured form.

Furthermore, we introduce the concept of a \textbf{self-evolving mechanism}. After each diagnostic cycle, the expert agent incrementally adjusts its internal rules based on feedback signals. These rules primarily govern repair suggestion generation, tool creation, and risk assessment. Through continuous evolution driven by historical experience, the expert agent progressively refines its decision-making process and becomes capable of handling increasingly complex configuration errors.

\begin{table*}[t]
  \centering
  \small
  \renewcommand{\arraystretch}{1.25}
  \begin{tabular}{l c c c c c}
    \noalign{\global\arrayrulewidth=1.2pt}\hline
    \noalign{\global\arrayrulewidth=0.6pt}

    \multirow{2}{*}{\textbf{Method}} 
    & \multirow{2}{*}{\textbf{Backbone}}
    & \multirow{2}{*}{\textbf{DGSR}} 
    & \textbf{\# Successfully} 
    & \multirow{2}{*}{\textbf{EBSR}} 
    & \textbf{\# Successfully} \\
    & 
    & 
    & \textbf{Generated Dockerfiles} 
    & 
    & \textbf{Built Environments} \\
    \hline

    pipreqs~\citep{pipreqs}        
      & -
      & 29.8\% & 125 & 6.0\%  & 25  \\

    LLM generator~\citep{hu2025reporun}  
      & GPT-4o
      & 47.6\% & 200 & 22.1\% & 93  \\

    SWE-agent~\citep{yang2024swe}      
      & GPT-4o
      & 26.9\% & 113 & 9.0\%  & 38  \\

    Repo2Run~\citep{hu2025reporun}       
      & GPT-4o
      & 100\%  & 420 & 86.0\% & 361 \\
    \hline

    \rowcolor{gray!15}
    \textbf{EvoConfig} 
      & GPT-4o
      & \textbf{100\%} 
      & \textbf{420} 
      & \textbf{88.1\%} 
      & \textbf{370} \\

    \noalign{\global\arrayrulewidth=1.2pt}\hline
  \end{tabular}
  \caption{
    Main results of different baselines in terms of Dockerfile generation
    and environment build success under the same backbone.
  }
  \label{tab:dgdr_ebsr_comparison}
\end{table*}

\begin{table}[t]
  \centering
  \small
  \setlength{\tabcolsep}{4pt}
  \begin{tabular}{p{1.4cm} l c c}
    \noalign{\global\arrayrulewidth=1.2pt}\hline
    \noalign{\global\arrayrulewidth=0.6pt}

    \multirow{2}{*}{\textbf{Method}} 
    & \multirow{2}{*}{\textbf{Backbone}} 
    & \multirow{2}{*}{\textbf{EBSR}} 
    & \rule{0pt}{2.2ex}\textbf{\# Successfully Built} \\
    & 
    & 
    & \textbf{Environments} \\
    \hline

    \multirow{2.5}{*}{Repo2Run}
      & \rule{0pt}{3.2ex}GPT-3.5-turbo     & 71.0\% & 230 \\
      & \rule{0pt}{3.2ex}GPT-4o-mini       & 40.0\% & 12 \\
    \hline

    \multirow{2.5}{*}{\textbf{EvoConfig}}
      & \rule{0pt}{3.2ex}GPT-3.5-turbo     & \cellcolor{gray!15}\textbf{78.1}\%

 & \cellcolor{gray!15}\textbf{253} \\
      & \rule{0pt}{3.2ex}GPT-4o-mini       & \textbf{46.7}\% 
  &\textbf{14} \\

    \noalign{\global\arrayrulewidth=1.2pt}\hline
  \end{tabular}
  \caption{
Performance comparison under different backbone models on the 324 repositories from EnvBench. Results for \texttt{gpt-4o-mini} are obtained on a randomly sampled
subset of 30 repositories.
  }
  \label{tab:backbone_single_column}
\end{table}

\section{Experimental Setup}

We evaluate EvoConfig from two complementary perspectives: environment build success and process-level error correction capability.

\subsection{Environment Build Success Evaluation}

\paragraph{Dataset and Baselines.}
We evaluate environment construction on 420 Repo2Run repositories~\citep{hu2025reporun} and 324 Python repositories from EnvBench~\citep{eliseeva2025envbench}, excluding 5 EnvBench repositories larger than 200MB. All experiments follow the Repo2Run protocol and compare EvoConfig with \textbf{pipreqs}~\citep{pipreqs}, \textbf{LLM Generator}, \textbf{SWE-agent}~\citep{yang2024swe}, and \textbf{Repo2Run}~\citep{hu2025reporun}. We use \texttt{gpt-4o-2024-05-13}, \texttt{GPT-3.5-turbo}, and \texttt{GPT-4o-mini}, with a 2-hour time limit and up to 100 interaction rounds. Additional details are provided in Appendix~\ref{sec:appendix-env_evaluation}.

\paragraph{Evaluation Metrics.}
We use two metrics to evaluate environment construction. \textbf{DGSR} measures the percentage of attempts that generate a runnable Dockerfile that builds without errors, while \textbf{EBSR} measures the percentage of attempts that successfully build executable environments, requiring both a successful Dockerfile build and the ability to execute tests with \texttt{pytest}, regardless of test outcomes.

\subsection{Process Error Correction Evaluation}

\paragraph{Dataset and Baselines.}
For process-level error correction evaluation, we use the EnConda-Bench dataset~\citep{kuang2025processleveltrajectoryevaluationenvironment}, which is designed to assess an agent’s ability to diagnose errors and recover from failed configuration steps during interactive execution. We evaluate all 4,201 instances provided by EnConda-Bench and compare EvoConfig against representative baselines, including \textbf{SWE-Agent}, \textbf{OpenHands}~\citep{wang2024openhands}, \textbf{INSTALLAMATIC}~\citep{milliken2025beyond}, and \textbf{Repo2Run}~\citep{hu2025reporun}, using \texttt{GPT-4.1} and \texttt{DeepSeek-V3}~\citep{liu2023repobench} as the underlying language models. More details about our selected baselines are provided in Appendix~\ref{sec:appendix-process_evaluation}.

\paragraph{Evaluation Metrics.}
We follow the evaluation protocol and metrics defined in EnConda-Bench, which measure an agent’s capability from error perception to corrective execution. Specifically, the metrics include error classification precision and recall, error description accuracy and fix accuracy. Each agent interacts with the execution environment step by step, generates diagnostic feedback and repair actions upon failure, and is evaluated based on both the correctness of intermediate error handling and the final recovery outcome.

\begin{table*}[t]
\centering
\small
\setlength{\tabcolsep}{5.5pt}
\renewcommand{\arraystretch}{1.15}
\begin{tabular}{l l c c c c c}
\toprule
\multirow{3}{*}{\centering\textbf{Method}} 
& \multirow{3}{*}{\centering\textbf{Backbone}} 
& \multicolumn{3}{c}{\textbf{Perception}} 
& \textbf{Feedback} 
& \textbf{Feedback and Action} \\
\cmidrule(lr){3-5}
\cmidrule(lr){6-6}
\cmidrule(lr){7-7}
& 
& \multicolumn{3}{c}{Error type} 
& Error description 
& Fix suggestion \\
\cmidrule(lr){3-5}
\cmidrule(lr){6-6}
\cmidrule(lr){7-7}
& 
& Pre. 
& Rec. 
& F1 
& ACC. 
& ACC. \\

    \noalign{\global\arrayrulewidth=0.6pt}\hline
    \noalign{\global\arrayrulewidth=0.6pt}

\multicolumn{7}{c}{\textit{Code Agent}} \\
\midrule
\multirow{2}{*}{SWE-Agent~\citep{yang2024swe}}
& GPT-4.1        & 43.7 & 83.2 & 55.3 & 49.8 & 30.7 \\
& DeepSeek-V3    & 41.2 & 70.3 & 51.9 & 44.5 & 27.8 \\
\midrule
\multirow{2}{*}{OpenHands~\citep{wang2024openhands}}
& GPT-4.1        & 42.5 & 72.0 & 53.2 & 46.0 & 29.1 \\
& DeepSeek-V3    & 46.7 & \textbf{93.6} & 58.7 & 51.9 & 33.8 \\
\midrule

\multicolumn{7}{c}{\textit{Environment Configuration Agent}} \\
\midrule
\multirow{2}{*}{INSTALLAMATIC~\citep{milliken2025beyond}}
& GPT-4.1        & 37.5 & 70.4 & 48.9 & 45.3 & 29.1 \\
& DeepSeek-V3    & 40.7 & 76.8 & 53.2 & 49.3 & 32.5 \\
\midrule
\multirow{2}{*}{Repo2Run~\citep{hu2025reporun}}
& GPT-4.1        & 44.2 & 72.3 & 54.8 & 48.5 & 38.6 \\
& DeepSeek-V3    & 46.3 & 74.2 & 56.8 & 44.6 & 41.2 \\
\midrule
\multirow{2}{*}{\textbf{EvoConfig}}
& GPT-4.1        
& \cellcolor{gray!15}49.2 
& \cellcolor{gray!15}75.4 
& \cellcolor{gray!15}59.7 
& \cellcolor{gray!15}\textbf{56.5} 
& \cellcolor{gray!15}39.4 \\
& DeepSeek-V3    
& \cellcolor{gray!15}\textbf{52.3} 
& \cellcolor{gray!15}77.9 
& \cellcolor{gray!15}\textbf{62.6} 
& \cellcolor{gray!15}48.3 
& \cellcolor{gray!15}\textbf{45.9} \\

\bottomrule
\end{tabular}

\caption{
Main results across different agents on EnConda-Bench.
}
\label{tab:agent_capability}
\end{table*}

\section{Result Analysis}

\subsection{Main Results}

\paragraph{Environment Construction Success Analysis.}

The results of different baselines are presented in Table \ref{tab:dgdr_ebsr_comparison}. Results of all baselines except EvoConfig are taken from the original Repo2Run benchmark to ensure a fair comparison.

We observe that EvoConfig achieves an environment building success rate that is comparable to, and slightly higher than, Repo2Run on the original set of 420 repositories. EvoConfig successfully builds executable environments for 370 repositories (EBSR 88.1\%), compared to 361 repositories (EBSR 86.0\%) built by Repo2Run. Given the already strong performance of Repo2Run, this improvement suggests that EvoConfig can recover a small but non-negligible fraction of failure cases that remain challenging for existing environment configuration agents. EvoConfig also maintains a DGSR of 100\%, matching Repo2Run and confirming that robust rollback and verification mechanisms are preserved, while other baselines fail to consistently guarantee Dockerfile buildability.

Table~\ref{tab:backbone_single_column} further presents environment building performance under different language model backbones. For \texttt{gpt-3.5-turbo}, EvoConfig improves EBSR from 71.0\% to 78.1\%, corresponding to 23 additional repositories successfully configured. We also present results using \texttt{gpt-4o-mini}, evaluated on a randomly sampled subset of 30 repositories due to computational constraints. This result indicates that the advantages of EvoConfig generalize across different model backbones.

\paragraph{Process-level Error Correction Analysis.} We evaluate process-level error correction results on EnConda-Bench in Table~\ref{tab:agent_capability}. EvoConfig demonstrates consistently stronger performance across both error perception and repair-related metrics, indicating improved handling of configuration failures during interactive execution.

We observe that code agents such as SWE-Agent and OpenHands show improved error perception compared to generic agents, but their ability to translate diagnosis into effective repair actions remains limited. For instance, OpenHands with DeepSeek-V3 achieves an error type F1 score of 58.7 and an error description accuracy of 51.9, while its fix suggestion accuracy is only 33.8, indicating a clear gap between error understanding and action-level repair. Environment configuration agents further improve repair effectiveness: INSTALLAMATIC increases fix accuracy to 32.5, and Repo2Run reaches 41.2 under DeepSeek-V3, demonstrating the benefit of explicitly modeling environment interaction. EvoConfig consistently achieves the strongest performance across both backbones, reaching error type F1 scores of 59.7/62.6 and fix suggestion accuracies of 39.4/45.9 under GPT-4.1 and DeepSeek-V3, respectively. These results suggest that EvoConfig better aligns fine-grained error analysis with actionable repair guidance, highlighting the value of adaptive, expert-driven diagnosis in improving process-level error correction.

\begin{table}[t]
  \centering
    \small
  \begin{tabular}{p{2.7cm}cc}
    \noalign{\global\arrayrulewidth=1.2pt}\hline
    \noalign{\global\arrayrulewidth=0.6pt}

    \multirow{2}{*}{\textbf{Method}} 
    & \multirow{2}{*}{\textbf{EBSR}} 
    & \textbf{\# Successfully} \\
    & 
    & \textbf{Built Environments} \\
    \hline

    {\raggedright
    \rule{0pt}{3.2ex}\textbf{w/o Environment \\ 
    Info Extraction}}
      & 82.0\%  & 82  \\

     \rule{0pt}{3.2ex}{\raggedright
    \textbf{w/o Self-Evolving\\
    Expert Diagnosis}}
      & 75.0\% & 75  \\

    \hline

    \rowcolor{gray!15}
    \rule{0pt}{3.2ex}\textbf{EvoConfig} 
      & 83.0\% 
      & 83 \\

    \noalign{\global\arrayrulewidth=1.2pt}\hline
  \end{tabular}
  \caption{
    Ablation results of EvoConfig in terms of environment build success.
  }
  \label{tab:ebsr_comparison}
\end{table}

\subsection{Ablation Study}

\begin{figure}[t]
    \centering
    \includegraphics[width=0.39\textwidth]{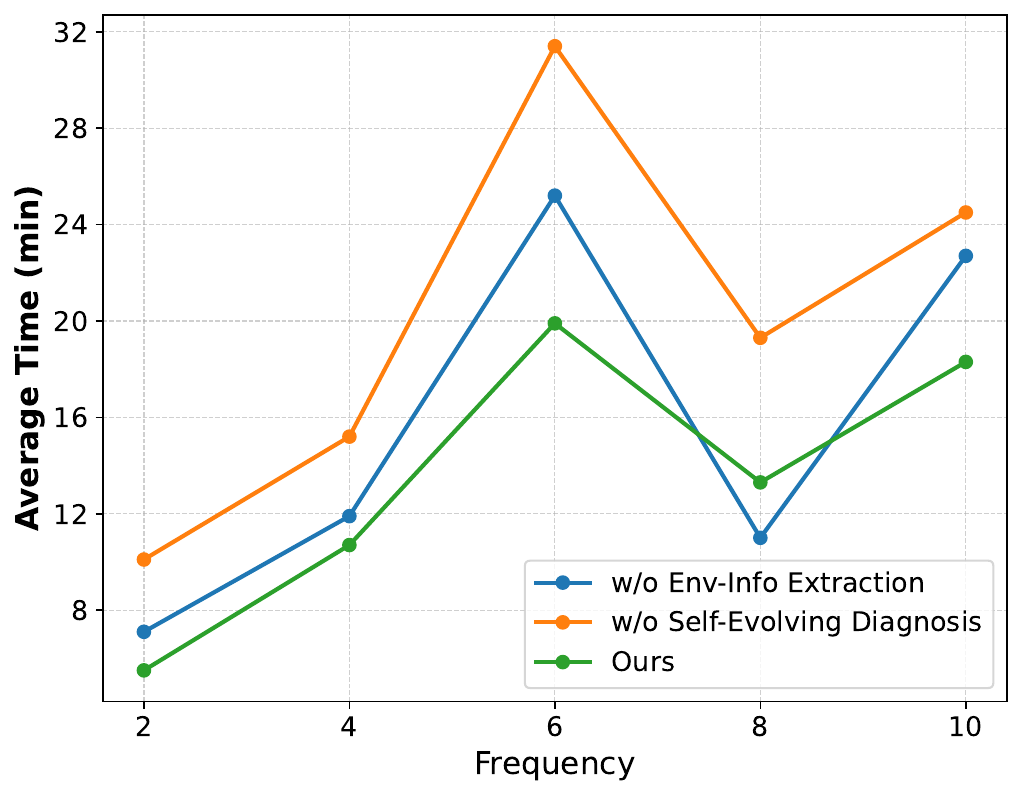}
    \caption{Runtime distribution of successful environment builds in the ablation study.}
    \label{fig:ablation}
\end{figure}

We conduct ablation studies on a randomly sampled set of 100 repositories from EnvBench to examine the contributions of the environment information extraction module and the self-evolving expert diagnosis module, with additional details provided in Appendix~\ref{sec:appendix-Time}. As shown in Table~\ref{tab:ebsr_comparison}, removing the self-evolving expert diagnosis module leads to a substantial drop in environment building success rate (EBSR) from 83.0\% to 75.0\%, while removing the environment information extraction module results in a smaller decrease to 82.0\%. The runtime comparison in Figure~\ref{fig:ablation}, measured on 30 repositories successfully configured by all variants, further shows that EvoConfig consistently achieves lower average configuration time. In particular, disabling environment information extraction leads to longer execution trajectories, whereas removing expert diagnosis causes the most significant slowdown and higher variance, indicating repeated and inefficient repair attempts. These results suggest that environment information extraction mainly improves efficiency, while adaptive diagnosis is critical for robustness and success.

\begin{figure}[t]
    \centering
    \includegraphics[width=0.39\textwidth]{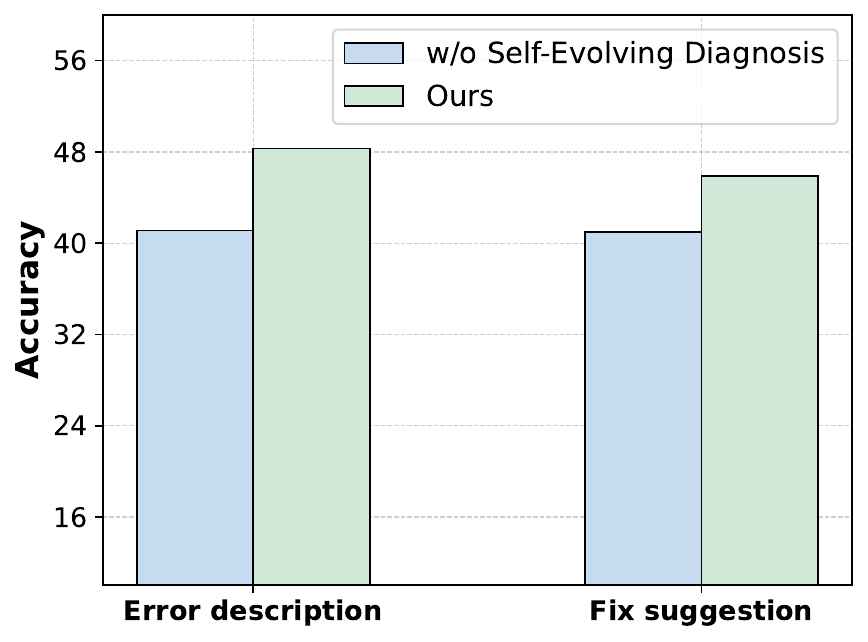}
    \caption{Ablation results of EvoConfig on process-level error correction.}
    \label{fig:ablation_error}
\end{figure}

To further evaluate the effectiveness of the self-evolving expert diagnosis module at the process level, we conduct a focused ablation study on EnConda-Bench using DeepSeek-V3, with results shown in Figure~\ref{fig:ablation_error}. The evaluation is performed on all instances from a randomly sampled set of 100 repositories, and more detailed analysis is provided in Appendix~\ref{sec:appendix-Expert}. Removing the diagnosis module consistently degrades performance across all stages of error handling, with the error description accuracy decreasing from 48.3 to 44.1, and fix suggestion accuracy declining from 45.9 to 41.0. Notably, the performance gap is most pronounced in fix suggestion accuracy, indicating that without adaptive expert feedback, the agent struggles to translate error understanding into effective corrective actions. In contrast, EvoConfig maintains a more consistent perception feedback action performance, indicating that the self evolving diagnostic mechanism enhances error correction ability of agent at the process-level while ensuring environment configuration success rate.

\begin{table}[t]
  \centering
  \small
  \setlength{\tabcolsep}{10pt}
  \begin{tabular}{l c c c}
    \noalign{\global\arrayrulewidth=1.2pt}\hline
    \noalign{\global\arrayrulewidth=0.6pt}

    \rule{0pt}{3.2ex}\textbf{Method} 
    & \textbf{Times (min)} 
    & \textbf{Tokens} 
    & \textbf{Cost} \\
    \hline

    \rule{0pt}{3.2ex}Repo2Run &30.5 & 495268 & \$0.33 \\
    \rule{0pt}{3.2ex}\textbf{EvoConfig}  & \textbf{20.9} & \textbf{229531} & \textbf{\$0.16} \\

    \noalign{\global\arrayrulewidth=1.2pt}\hline
  \end{tabular}
  \caption{
    Effiency comparison of successful building.
  }
  \label{tab:token_cost}
\end{table}

\section{Discussion}

\subsection{Efficiency and Cost Analysis}

As shown in Table~\ref{tab:token_cost}, experiments on the 324 EnvBench repositories show that EvoConfig substantially improves both aspects: it reduces the average configuration time per repository from 30.5 minutes to 20.9 minutes and incurs lower token-level and monetary cost under \texttt{gpt-3.5-turbo}. These gains are largely attributable to EvoConfig’s multi-agent design, which separates execution control from error diagnosis and feedback interpretation, preventing long execution traces from repeatedly entering the main agent’s context and thereby reducing redundant reasoning during configuration.

\subsection{Failure Case Study}

We analyze the failure cases of EvoConfig on the EnvBench benchmark, with the distribution summarized in Table~\ref{tab:failure_category}. Most failures are caused by external execution constraints or repository-intrinsic issues rather than limitations of the agent itself. Hardware insufficiency is the most common failure source, accounting for 32.4\% of failed cases, followed by missing or incomplete configuration information (28.2\%), where repositories lack core files such as \texttt{pyproject.toml}, \texttt{setup.py}, or \texttt{requirements.txt}. A further portion of failures arises from execution timeouts during dependency installation or test execution, reflecting practical limits imposed by heavy dependencies and long-running tests.

\begin{table}[t]
  \centering
  \small
  \setlength{\tabcolsep}{10pt}
  \renewcommand{\arraystretch}{1.15}
  \begin{tabular}{l r}
    \noalign{\global\arrayrulewidth=1.2pt}\hline
    \rowcolor{gray!15}
    \textbf{Category} & \textbf{\# Case (\%)} \\
    \noalign{\global\arrayrulewidth=0.6pt}\hline
    Hardware Insufficiency          & 23 (32.4\%) \\
    Config Files Missing                   & 20  (28.2\%)  \\
   Dependency Installation Timeout              & 10  (14.1\%)  \\
    Unit Tests Missing
 & 5  (7.0\%) \\
    Runtest Timeout                 & 13 (18.3\%) \\
    \noalign{\global\arrayrulewidth=1.2pt}\hline
  \end{tabular}
  \caption{
    Analysis of failure cases in EvoConfig.
  }
  \label{tab:failure_category}
\end{table}

\section{Related Work}
\label{sec:related-work}

\paragraph{Executable environments as a prerequisite for training and evaluating SWE agents.}
Executable environments are a prerequisite for repository-level SWE agents, because both training signals and evaluation protocols assume that projects can be built and their verification procedures can be executed reproducibly. Accordingly, environment configuration is deeply embedded in popular benchmarks and data pipelines: several widely-used settings rely on manual, repository-specific environment curation, such as SWE-bench~\citep{jimenez2024swebench}, SWE-Flow~\citep{zhang2025synthesizing}, SWE-Gym~\citep{pan2025training}, and R2E-Gym~\citep{jain2025regym}. Recent benchmark and data construction workflows increasingly incorporate automated or semi-automated environment synthesis as a critical stage, including SetupAgent~\citep{vergopoulos2025automated}, SWE-smith~\citep{yang2025swesmith}, SWE-Factory~\citep{guo2025swefactoryautomatedfactoryissue}, SWE-bench-Live~\citep{zhang2025swebench}, SWE-Compass~\citep{xu2025swecompassunifiedevaluationagentic}, and SWE-Bench++~\citep{wang2025swebenchframeworkscalablegeneration}. Collectively, these trends motivate environment synthesis as a first-class research problem that directly controls the scalability and reliability of executable SWE data.

\paragraph{Methods for automated environment setup.}
Automated setup methods broadly fall into \emph{deterministic} and \emph{agentic} families~\cite{li2024benchmarking, li2025towards,li2025refine,kuang2025natural, ye2025productagent}. Deterministic approaches implement setup as automated scripts or fixed pipelines, which execute standardized procedures across diverse repositories to maximize reproducibility and reduce per-repository manual effort; a representative example is R2E~\citep{jain2024re}, which instantiates executable test environments via scripted setup procedures. EnvBench~\citep{eliseeva2025envbench} spans both families: it introduces a benchmark for automated environment setup, includes a deterministic shell-script baseline, and also evaluates LLM-based Bash agents under the same task definition and proxy-based verification signals. Template-guided container synthesis constrains Dockerfile structure while leaving repository-specific slots to be filled, improving robustness at scale in SWE-Bench++~\citep{wang2025swebenchframeworkscalablegeneration}. Agentic approaches treat setup as interactive search-and-repair: an LLM agent retrieves commands from documentation and project artifacts, executes them in a sandbox, diagnoses failures from logs, and iteratively refines the plan, as done in SetupAgent~\citep{vergopoulos2025automated} and in the RepoLaunch pipeline of SWE-bench-Live~\citep{zhang2025swebench}; SWE-Factory adopts multi-agent decomposition and environment reuse to amortize successful configurations~\citep{guo2025swefactoryautomatedfactoryissue}. Related systems target key subroutines, including scalable export of runnable Docker environments in Repo2Run~\citep{hu2025reporun}, installation under incomplete documentation in Installamatic~\citep{DBLP:conf/saner/MillikenKY25}, and test execution for arbitrary projects in ExecutionAgent~\citep{DBLP:journals/pacmse/BouzeniaP25}, while earlier dependency inference in DockerizeMe illustrates the limits of purely static signals~\citep{DBLP:conf/icse/HortonP19}.

\paragraph{Benchmarks that evaluate environment setup ability.}
With the rapid development in the field of LLMs~\citep{li2025admtree, chen2025dast, huang2024lateval, li2024llms, zhang2025cold,liu2022we,du2024llms,li2022past,li2023effectiveness, li2025rethinking, xu2025let, yu2024seqgpt}, a complementary line of work elevates environment bootstrapping into a first-class benchmarked capability~\citep{miao2025recode,li2025one,chen2025coderankeval}. EnvBench~\citep{eliseeva2025envbench} provides a large-scale benchmark for repository-specific setup across Python and JVM projects and introduces automatic proxy metrics such as missing-import and compilation checks to support scalable evaluation. SetupBench~\citep{arora2025setupbenchassessingsoftwareengineering} formalizes bootstrapping from a bare Linux sandbox with deterministic one-line verification commands, enabling fine-grained analysis of failure modes such as incomplete toolchains and non-persistent modifications. Enconda-bench~\citep{kuang2025processleveltrajectoryevaluationenvironment} moves beyond end-to-end success by scoring \emph{process-level} trajectories and diagnosing capabilities such as setup planning, error localization, and feedback-driven repair under realistically perturbed instructions. SWE-Compass~\citep{xu2025swecompassunifiedevaluationagentic} incorporates configuration and deployment tasks into a broader agentic coding evaluation suite, contextualizing setup as part of end-to-end agent behavior. Finally, Multi-Docker-Eval~\citep{fu2025multidockerevalshovelgoldrush} expands evaluation to multi-language repositories and emphasizes both effectiveness and efficiency, including time and resource usage as well as resulting image size.

\section{Conclusion}

In this paper, we propose EvoConfig, a self-evolving multi-agent framework that decouples execution, diagnosis, and repair. By combining lightweight environment information extraction with adaptive expert diagnosis, EvoConfig improves configuration robustness and efficiency. Experiments on multiple benchmarks demonstrate strong environment building performance with reduced time and token cost, while process-level evaluations show improved error understanding and repair quality. EvoConfig focuses on enabling test execution, and extending it to reason about test outcomes remains future work.

\section*{Limitations}

EvoConfig focuses on constructing executable environments and improving process-level error correction, but does not reason about test correctness. Our evaluation considers whether unit tests can be executed, rather than the proportion of tests that pass. In practice, test failures may arise from issues beyond environment configuration. While EvoConfig enables tests to run reliably, analyzing test outcomes and debugging failing tests remain outside its current scope and are left for future work.

\section*{Ethical Considerations}
\paragraph{Potential Risks} Although EvoConfig improves robustness and efficiency in automated environment configuration, several risks remain. The framework depends on execution feedback quality, and noisy or incomplete errors may still affect diagnosis. In addition, the self-evolving mechanism may require sufficient feedback to stabilize in early stages. Finally, EvoConfig focuses on building runnable environments and does not guarantee the correctness of test outcomes, which may limit its use in strict functional validation settings.

\paragraph{Ethical Statement} This work focuses on automated environment configuration for open-source software repositories using large language model–based agents. All experiments are conducted on publicly available data and executed in isolated environments, without involving personal, sensitive, or private information.

\paragraph{LLMs Usage Statement} Large language models were used to assist with language polishing and clarity improvement during the writing process; all technical content, experimental design, and conclusions were developed and verified by the authors. The proposed method aims to improve research reproducibility and scalability in software engineering and does not introduce new ethical risks beyond those of existing automated development tools.


\bibliography{custom}

\appendix

\section{Environment Build Success Evaluation}
\label{sec:appendix-env_evaluation}

\subsection{Baselines}
We evaluate environment configuration performance using the following representative baselines:
\begin{itemize}
  \item \textbf{pipreqs}: A static dependency analysis tool that inspects Python import statements to infer required packages and generates a \texttt{requirements.txt} file, which is then used to construct a Dockerfile.
  \item \textbf{LLM Generator}: A direct LLM-based approach that parses repository \texttt{README} files and generates executable Dockerfiles without iterative interaction.
  \item \textbf{SWE-agent}: An LLM-based agent with a custom agent--computer interface that supports file inspection, editing, and command execution; its framework is retained while prompts are adapted for environment configuration.
  \item \textbf{Repo2Run}: A strong agent-based baseline specifically designed for iterative environment configuration through interaction with the execution environment.
\end{itemize}

\begin{table}[t]
  \centering
    \small
  \begin{tabular}{p{2.7cm}cc}
    \noalign{\global\arrayrulewidth=1.2pt}\hline
    \noalign{\global\arrayrulewidth=0.6pt}

    \multirow{1}{*}{\textbf{Method}} 
    & \multirow{1}{*}{\textbf{EBSR}} 
    & \textbf{Average Times (min)} \\
    \hline

    {\raggedright
    \rule{0pt}{3.2ex}\textbf{w/o Environment Info Extraction}}
      & 82.0\%  & 22.5  \\

     \rule{0pt}{3.2ex}{\raggedright
    \textbf{w/o Self-Evolving\\
    Expert Diagnosis}}
      & 75.0\% & 27.2  \\

    \hline

    \rowcolor{gray!15}
    \rule{0pt}{3.2ex}\textbf{EvoConfig} 
      & 83.0\% 
      & 20.5 \\

    \noalign{\global\arrayrulewidth=1.2pt}\hline
  \end{tabular}
  \caption{
    Ablation results of EvoConfig in terms of environment build success and time cost.
  }
  \label{appendix:ebsr_time}
\end{table}

\begin{table*}[t]
\centering
\small
\setlength{\tabcolsep}{6pt}
\renewcommand{\arraystretch}{1.25}
\begin{tabular}{c c c c c c c}
\toprule
\multirow{3}{*}{\centering\textbf{Method}} 
& \multirow{3}{*}{\centering\textbf{Backbone}} 
& \multicolumn{3}{c}{\textbf{Perception}}
& \textbf{Feedback}
& \textbf{Feedback and Action} \\
\cmidrule(lr){3-5}
\cmidrule(lr){6-6}
\cmidrule(lr){7-7}
& 
& \multicolumn{3}{c}{Error Type}
& {Error Description}
& {Fix Suggestion} \\
\cmidrule(lr){3-5}
\cmidrule(lr){6-6}
\cmidrule(lr){7-7}
& 
& \textbf{Pre.} 
& \textbf{Rec.} 
& \textbf{F1} 
& \textbf{ACC.} 
& \textbf{ACC.} \\
\midrule

\textbf{w/o Self-Evolving Expert Diagnosis}
& DeepSeek-V3
& 43.1 & 76.2 & 55.1 & 44.1 & 41.0 \\

\textbf{EvoConfig}
& DeepSeek-V3
&  \cellcolor{gray!15}\textbf{52.3} 
& \cellcolor{gray!15}\textbf{77.9} 
& \cellcolor{gray!15}\textbf{62.6} 
& \cellcolor{gray!15}\textbf{48.3} 
& \cellcolor{gray!15}\textbf{45.9} \\

\bottomrule
\end{tabular}

  \caption{
    Complete ablation experiment results of EvoConfig on process-level error correction.
  }
  \label{appendix:complete}
  
\end{table*}

\section{Process Error Correction Evaluation}
\label{sec:appendix-process_evaluation}

To evaluate process-level error correction, we compare EvoConfig with several representative agent-based baselines that support iterative interaction with the execution environment.

\begin{itemize}
  \item \textbf{SWE-agent}: An LLM-based agent originally designed for automated bug fixing, which supports file inspection, code editing, and command execution through a custom agent--computer interface. We adapt its prompts for process-level environment error correction.
  \item \textbf{OpenHands}: A general-purpose autonomous agent framework for software engineering tasks, used here as a generic baseline to assess its ability to correct environment errors through multi-step interaction.
  \item \textbf{INSTALLAMATIC}: An LLM-driven system that focuses on generating installation and setup commands for resolving dependency-related environment issues, without explicit long-horizon agent planning.
  \item \textbf{Repo2Run}: A specialized agent-based system for repository environment configuration.
\end{itemize}

\section{Time Efficiency Analysis}
\label{sec:appendix-Time}
We analyze environment building success rates (EBSR) and time efficiency of different variants, as reported in Table~\ref{appendix:ebsr_time}. EvoConfig achieves both the highest success rate (83.0\%) and the lowest average configuration time (20.5 minutes), indicating that its improved performance does not come at the cost of increased runtime. In contrast, removing the environment information extraction module slightly reduces the success rate to 82.0\% while increasing the average configuration time to 22.5 minutes, suggesting that limited environment awareness leads to inefficient trial-and-error and correspondingly longer execution trajectories.

Specifically, the success rate drops substantially to 75.0\%, accompanied by a significant increase in average configuration time to 27.2 minutes. This observation indicates that static diagnosis strategies not only reduce the likelihood of successful environment construction but also result in repeated and inefficient repair attempts, thereby prolonging the overall configuration process. Together, these results demonstrate that while environment information extraction mainly contributes to execution efficiency, adaptive expert diagnosis is crucial for achieving both high configuration success rates and low configuration time.

\section{Effect of Self-Evolving Expert Diagnosis}
\label{sec:appendix-Expert}
We evaluate the contribution of the self-evolving expert diagnosis module by comparing the full EvoConfig framework with a variant that removes this component while using the same backbone (DeepSeek-V3). As shown in Table~\ref{appendix:complete}, removing self-evolving diagnosis results in a consistent performance degradation across all evaluation stages, indicating its critical role in the overall system.

From the perception perspective, the absence of self-evolving diagnosis leads to a noticeable drop in error type recognition performance, with the F1 score decreasing from 62.6 to 55.1, mainly due to reduced precision. This suggests that static expert behavior limits the agent’s ability to accurately identify error patterns. Moreover, the accuracy of error description generation also declines from 48.3\% to 44.1\%, reflecting less precise feedback when adaptive diagnosis is disabled.

The impact is further reflected in the action stage, where fix suggestion accuracy drops from 45.9\% to 41.0\%. Since effective repair actions rely on accurate error understanding, these results demonstrate that self-evolving expert diagnosis is an essential component for maintaining coherent perception--feedback--action alignment in large-scale environment configuration.

\end{document}